\documentclass[review]{elsarticle}

\usepackage{lineno,hyperref}
\usepackage{moreverb}
\usepackage{graphicx, wrapfig}
\usepackage{algorithm}
\usepackage{algpseudocode}
\usepackage{color}
\usepackage{amsmath}
\usepackage{txfonts}
\usepackage{afterpage}
\usepackage[margin=1in]{geometry}
\usepackage{mathrsfs}
\usepackage{tikz}
\usepackage{pstricks}
\usepackage{diagbox}
\usepackage{arydshln}
\usepackage{dsfont}
\usepackage{smartdiagram}
\usepackage{todonotes}
\usepackage{verbatim}
\usepackage{smartdiagram}
\usepackage{bm}
\usepackage{commath}
\usepackage{bbm}
\usepackage{mathrsfs}
\usepackage{dsfont}
\usepackage{extarrows}
\usepackage{subfigure}
\usepackage{multirow}
\usepackage{graphicx}
\usepackage{array}
\usepackage{epstopdf}
\usepackage{float}
\usepackage{booktabs}
  
\usetikzlibrary{calc,trees,positioning,arrows,chains,shapes.geometric,%
    decorations.pathreplacing,decorations.pathmorphing,shapes,%
    matrix,shapes.symbols}

\tikzset{
>=stealth',
  punktchain/.style={
    rectangle, 
    rounded corners, 
    draw=black, very thick,
    text width=10em, 
    minimum height=2em, 
    text centered, 
    on chain},
  line/.style={draw, thick, <-},
  element/.style={
    tape,
    top color=white,
    bottom color=blue!50!black!60!,
    minimum width=6em,
    draw=blue!40!black!90, very thick,
    text width=10em, 
    minimum height=3.5em, 
    text centered, 
    on chain},
  every join/.style={->, thick,shorten >=1pt},
  decoration={brace},
  tuborg/.style={decorate},
  tubnode/.style={midway, right=2pt},
}

\tikzstyle{arrow} = [thick,->,>=stealth]
\usepackage{lineno}
\usepackage[T1]{fontenc}

\newdefinition{rmk}{Remark}
\newproof{pf}{Proof}
\newcommand{\Neless}{\textbf{Ne80}}
\newcommand{\Nemore}{\textbf{Ne150}}
\usepackage{booktabs}
\usepackage{siunitx}

\newcommand{\gp}{\mathcal{GP}}
\newcommand{\bxi}{\bm{\xi}}

\newcommand{\comm}[1]{}

\makeatletter
\def\ps@pprintTitle{%
   \let\@oddhead\@empty
   \let\@evenhead\@empty
   \let\@oddfoot\@empty
   \let\@evenfoot\@oddfoot
}
\makeatother

\modulolinenumbers[1]

\journal{Journal of Computational Physics}
\let\oldequation\equation
\let\oldendequation\endequation

\renewenvironment{equation}
  {\linenomathNonumbers\oldequation}
  {\oldendequation\endlinenomath}

\let\oldalign\align
\let\oldendalign\endalign

\renewenvironment{align}
  {\linenomathNonumbers\oldalign}
  {\oldendalign\endlinenomath}

\begin{document}
\begin{frontmatter}
		
\title{An ILUES-based adaptive Gaussian process method for multimodal Bayesian inverse problems}
		
		
\author[UHaddress]{Zhihang Xu\fnref{co-first}}
\ead{zxu29@central.uh.edu}
\fntext[co-first]{These authors contributed equally to this work.}
\address[UHaddress]{Department of Mathematics, University of Houston, Houston 77204, USA}
\author[ShanghaiTechAddress]{Xiaoyu Zhu\fnref{co-first}}
\ead{zhuxiaoyu@sinopeda.com}
\address[ShanghaiTechAddress]{School of Information Science and Technology, ShanghaiTech University, Shanghai 201210, China}

\author[DaojiLiAddress]{Daoji Li}
\ead{dali@fullerton.edu}

\author[ShanghaiTechAddress]{Qifeng Liao\corref{mycorrespondingauthor}}
		\cortext[mycorrespondingauthor]{Corresponding author}
		\ead{liaoqf@shanghaitech.edu.cn}

\address[DaojiLiAddress]{Department of Information Systems and Decision Sciences, California State University, Fullerton 92831, USA}

\begin{abstract}
Inverse problems are prevalent in both scientific research and engineering applications. In the context of Bayesian inverse problems, sampling from the posterior distribution can be particularly challenging when the forward models are computationally expensive. This challenge is further compounded when the posterior distribution is multimodal. To address this issue, we propose a Gaussian process (GP)-based method to indirectly build surrogates for the forward model. Specifically, the unnormalized posterior density is expressed as a product of an auxiliary density and an exponential GP surrogate. Iteratively, the auxiliary density converges to the posterior distribution, starting from an arbitrary initial density. However, the efficiency of GP regression is highly influenced by the quality of the training data. Therefore, we utilize the iterative local updating ensemble smoother (ILUES) to generate high-quality samples that are concentrated in regions with high posterior probability. Subsequently, based on the surrogate model and mode information extracted using a clustering method, Markov chain Monte Carlo (MCMC) with a Gaussian mixed (GM) proposal is used to draw samples from the auxiliary density. Through numerical examples, we demonstrate that the proposed method can accurately and efficiently represent the posterior with a limited number of forward simulations.
\end{abstract}

\begin{keyword}
Bayesian inverse problems, Multimodal, Gaussian process, Surrogate, ILUES
\end{keyword}

\end{frontmatter}

\section{Introduction}\label{sec:intro}



Inverse problems have 
a wide range of applications
across many fields in science and engineering, such as weather prediction, groundwater flows, medicine, and molecular 
dynamics~\citep{tarantola2005inverse,kruschke2010bayesian,asch2016data,rizzi2011bayesian}. 
Inverse problems typically involve learning the inputs to a mathematical model, such as physical parameters and initial conditions, based on partial and noisy observations of model outputs.
Bayesian inverse problems (BIPs) provide a rigorous foundation to quantify uncertainty in the inverse solution~\citep{kaipio2005statistical}.
BIPs capture the uncertainty associated with parameters by 
incorporating prior knowledge about the model inputs into a probability distribution, known as the prior distribution.
By conditioning the prior distribution on the observed data, a posterior distribution is generated, providing a more accurate representation of the model inputs.
However, forward models, which map the parameters to measurable data in inverse problems, 
are often computationally expensive and complex when involving partial differential equations. 

In many applications, the posterior distribution does not have a closed-form expression and must be approximated. In such cases, Markov Chain Monte Carlo (MCMC) is commonly used as a powerful method for exploring 
the posterior distribution~\citep{andrieu2003introduction,cui2016dimension, beskos2022mcmc, dai2023bayesian, xu2024domainKL,xu2024domainVAE}.
The standard MCMC method consists of three 
steps in each iteration: (i) generating a candidate sample from the proposal distribution, (ii) computing the acceptance probability, and (iii) either accepting or rejecting the candidate sample. 
Through repeated iterations, the distribution of the generated samples ultimately converges to the posterior distribution.
However, when forward model evaluations are costly and the posterior distribution is multimodal, two main challenges arise for the standard MCMC approach. 
First, due to 
the Markov chain navigating low-probability barriers between modes, the MCMC sampler frequently becomes trapped within a single mode. Second, since MCMC methods require solving the forward model for every sample, 
the computational cost of MCMC simulations can become prohibitively high.
To address these challenges, several variants of the 
standard MCMC approach have been proposed to enhance its performance in handling multimodal posterior distributions. One notable variant is parallel tempering (PT), also known as replica exchange Monte Carlo, which 
runs multiple Markov chains in parallel with each chain exploring the target distribution at a different ``temperature'' and swaps states between any two chains at 
intervals~\citep{latz2021generalized, lin2022multi, lkacki2016state}. 
Another variant is the Differential Evolution Adaptive Metropolis (DREAM) algorithm, which combines the strengths of differential evolution and the Metropolis-Hastings algorithm. It has been widely used in inverse problems involving multimodal distributions and high-dimensional inputs~\citep{vrugt2009accelerating,vrugt2008treatment,zhang2020improving}. 
The key idea behind these variants for sampling multimodal distributions is to sufficiently explore the parameter space. However, these variants typically require more simulations than the standard MCMC.
When dealing with a computationally expensive forward model, an effective strategy is to develop a computationally inexpensive surrogate model, which can replace the original forward model for tasks such as calculating acceptance probabilities, parameter estimation, and uncertainty quantification.
Some popular surrogate models in BIPs include Gaussian process regression
~\citep{bilionis2013multi,seeger2004gaussian,wang2018adaptive}, 
polynomial chaos expansion~\citep{marzouk2007stochastic,li2014adaptive,zhang2020surrogate},  deep neural networks ~\citep{xia2022bayesian,zhu2018bayesian}, 
and local approximation~\citep{davis2022rate,conrad2016accelerating,conrad2018parallel}.
These surrogate models approximate the true forward model by using a training set of input-output pairs.
A good strategy for selecting the training set is to choose data points
from the high-density regions of the posterior 
distribution~\cite{sacks1989design,mackay1992information}. 
However, 
identifying the high-density regions of a multimodal posterior distribution is a challenging task. Moreover, for BIPs with multimodal posterior distributions, the combination of MCMC sampling and surrogate methods can 
can lead to unsatisfactory results.


The ensemble Kalman filter (EnKF), widely used in dynamical systems, is a derivative-free optimizer that selects samples to approximate a target distribution, which is typically assumed to be Gaussian.
The ensemble smoother (ES), a variant of EnKF, 
updates the samples utilizing all available data, leading to reduced computational costs~\citep{van1996data, skjervheim2011ensemble,emerick2013ensemble}. 
Both EnKF and ES can provide reliable estimates of the parameter of interest when a sufficiently large ensemble size is used.
However, due to the computationally intensive nature of the forward model, only a small ensemble is feasible.
Furthermore, EnKF and ES inherently provide a unimodal Gaussian approximation to the posterior distribution, which may fail to capture the complexities of the true distribution, particularly when the posterior is multimodal.
To address these issues, the iterative local updating ensemble smoother (ILUES) was proposed.  Its main idea is to iteratively update localized regions within the ensemble~\citep{zhang2018iterative}. 
One of the appealing properties of ILUES is its ability to concentrate the ensemble samples in regions with high posterior density.

In this paper, we propose an adaptive approximation sampler for multimodal posterior distributions by combining the adaptive Gaussian process (GP) framework and the ILUES method.
Specifically, the proposed approach consists of three main steps. 
First, we express the 
unnormalized posterior density as the product of an auxiliary density and the exponential of the target function, which is approximated using a 
GP surrogate. The training data of the GP surrogate are selected using the ILUES method since ILUES-generated samples can quickly accumulate in regions of high posterior density.
Next, we divide these ILUES-generated samples 
into $K$ clusters using the $K$-means method and obtain preliminary cluster information, including the number of clusters, as well as the mean and covariance of each cluster. Finally, we sample from the approximated posterior distribution using MCMC, with the proposal distribution being a Gaussian mixture whose parameters are derived from the final iteration of the ILUES method.

%
The rest of the paper is organized as follows. Section~\ref{sec2} introduces the formulation of Bayesian inverse problems. Section~\ref{sec 3} provides a brief review of the adaptive Gaussian process surrogate and the ILUES method. Section~\ref{sec:main_alg} presents the proposed approach for approximating multimodal posterior distributions. Section~\ref{sec 5} demonstrates the performance of the proposed method through numerical examples. Section~\ref{sec:con} concludes the paper.
\section{Bayesian inverse problems} \label{sec2}

In a Bayesian inverse problem, we seek to estimate a parameter vector $\theta \in \Theta \subseteq \mathbb{R}^{\bar{M}}$ based on observed data $\boldsymbol{d}_{obs} \in \mathbb{R}^{\bar{D}}$, where $\bar{M}$ is the dimension of the parameter space and $\bar{D}$ is the dimension of the data space. 
The forward model is typically defined as a function $\mathcal{G}:\mathbb{R}^{\bar{M}} \rightarrow{\mathbb{R}^{\bar{D}}}$. 
The 
relationship between 
$\theta$ and  $\boldsymbol{d}_{obs}$ is described by 
\begin{equation*}
\boldsymbol{d}_{obs} = \mathcal{G}(\theta) + \eta,
\end{equation*}
where $\eta$ represents the measurement noise.
Using Bayes' theorem, the posterior probability density of $\theta$ given $\boldsymbol{d}_{obs}$ is given by
\begin{equation*}
\boldsymbol{\pi}(\theta \vert \boldsymbol{d}_{obs}) = \frac{\mathcal{L}(\boldsymbol{d}_{obs} \vert \theta) \boldsymbol{\pi}_{prior}(\theta)}{\int \mathcal{L}(\boldsymbol{d}_{obs} \vert \theta)\boldsymbol{\pi}_{prior}(\theta) d\theta} \propto \mathcal{L}(\boldsymbol{d}_{obs}\vert \theta) \boldsymbol{\pi}_{prior}(\theta)\,.
\end{equation*}
Here $\mathcal{L}(\boldsymbol{d}_{obs} \vert \theta)$ is the likelihood and $\boldsymbol{\pi}_{prior}(\theta)$ is the prior, which reflects the initial knowledge of $\theta$ before observing the data. 
Under the assumption that $\eta$ is a $\bar{D}$-dimensional  zero mean Gaussian noise with the  non-singular covariance matrix $\boldsymbol{\Sigma}_\eta$, i.e., $\eta\sim\mathcal{N}(\boldsymbol{0},\boldsymbol{\Sigma}_\eta)$, the 
likelihood takes the following form
\begin{equation*}
\mathcal{L}(\boldsymbol{d}_{obs} \vert \theta) \propto \exp\left\{-2^{-1}\left[\boldsymbol{d}_{obs} - \mathcal{G}(\theta)\right]^T \boldsymbol{\Sigma}_\eta^{-1} \left[\boldsymbol{d}_{obs} - \mathcal{G}(\theta)\right]\right\}.
\end{equation*}
To get the uncertainty quantification of $\theta$, a commonly used approach  is to generate samples from the posterior distribution $\boldsymbol{\pi}(\theta \vert \boldsymbol{d}_{obs})$.
However, 
when the forward model $\mathcal{G}(\theta)$ is nonlinear in $\theta$, the posterior $\boldsymbol{\pi}(\theta \vert \boldsymbol{d}_{obs})$  
does not 
admit a closed-form expression.

\section{Overview of the adaptive GP surrogate and the ILUES method} \label{sec 3}

\subsection{Adaptive GP surrogate} \label{sec:adaptive_gp}

Computing the log-likelihood function $\mathcal{L}(\boldsymbol{d}_{obs} \vert \theta)$ is often computationally expensive, as it requires repeated evaluations of the forward model.
To address this challenge, a common approach is to employ
Gaussian process  regression to approximate the log-likelihood function. 
Let us briefly review the idea.
Let the target function be $F(\theta)$, and suppose we have access to $N$ data pairs, $\{(\theta_i, F(\theta_i))\}_{i=1}^N$.
The GP surrogate model predicts $F(\theta^*)$,
which is the value of the target function at any new point $\theta^*$. The key idea of Gaussian process regression is to assume that the target function $F(\theta)$ is a realization 
of a Gaussian random field, characterized by a mean function $\mu(\theta)$ and a covariance function defined by a kernel $k(\theta,\,\theta')$, 
i.e.,
$F(\theta)\sim \gp(\mu(\theta), k(\theta,\,\theta'))$. Write $\boldsymbol{x}
= [\theta_1, \ldots, \theta_N]$ and $\boldsymbol{y}
= [F(\theta_1),\ldots,F(\theta_N)]$.
Then 
the joint distribution of $\boldsymbol{y}$ and $F(\theta^*)$ is 
\begin{equation}
\label{eq Gaussian distribution} 
    \begin{bmatrix}
        \boldsymbol{y}\\
        F(\theta^*)\\
    \end{bmatrix}
 \sim \mathcal{N}
 \left(\begin{bmatrix} 
        \mu(\boldsymbol{x})\\
        \mu(\theta^*)\\
    \end{bmatrix},  \begin{bmatrix} K(\boldsymbol{x},\boldsymbol{x}) & K(\boldsymbol{x},\theta^*) \\ K(\theta^*,\boldsymbol{x}) & k(\theta^*,\theta^*) \end{bmatrix}\right),
\end{equation}
where 
$\mu(\boldsymbol{x})$ is an $N \times 1$ vector with the $i$-th element $\mu(\theta_i)$, $K(\boldsymbol{x},\boldsymbol{x})$ is an $N \times N$ covariance matrix with the $(i, j)$ entry 
$K_{ij} = k(\theta_i,\theta_j)$,   
$ K(\boldsymbol{x},\theta^*)$ is an $N \times 1$ vector with the $i$-th element 
$ k(\theta_i, \theta^*)$, and $ K(\theta^*,\boldsymbol{x})$ is the transpose of $ K(\boldsymbol{x},\theta^*)$.
It follows from~\eqref{eq Gaussian distribution} that the posterior distribution of $F(\theta^*)$ is given by
\begin{equation*} 
\boldsymbol{\pi}(F(\theta^*) \vert \boldsymbol{x},\boldsymbol{y},\theta^*)  = \mathcal{N}(\tilde{\mu}(\theta^*),\text{var}(F(\theta^*))),
\end{equation*}
where the posterior mean and variance are 
\begin{equation} \label{mu}
    \tilde{\mu}(\theta^*) = \mu(\theta^*) + K(\theta^*,\boldsymbol{x})[K(\boldsymbol{x},\boldsymbol{x})]^{-1}[\boldsymbol{y} - \mu(\boldsymbol{x})],
\end{equation}
and 
\begin{equation}\label{var}
    \text{var}(F(\theta^*)) = k(\theta^*,\theta^*) - K(\theta^*,\boldsymbol{x})[K(\boldsymbol{x},\boldsymbol{x})]^{-1}K(\boldsymbol{x},\theta^*),
\end{equation}
respectively. 
Thus, we can use~\eqref{mu} to approximate $F(\theta^*)$ and~\eqref{var} to quantify the approximation error, respectively.

However, when the log-likelihood function $\log \mathcal{L}(\boldsymbol{d}_{obs} \vert \theta)$ is highly nonlinear and complex, 
this approach may not provide accurate approximations~\citep{bilionis2013multi}.
To address this 
issue, \citep{wang2018adaptive} proposed an adaptive Gaussian process scheme. 
%
By Bayes' theorem, the unnormalized posterior probability density can be expressed as
\begin{equation} \label{unnormalized posterior}
\boldsymbol{\tilde{\pi}}(\theta \vert \boldsymbol{d}_{obs}) = \mathcal{L}(\boldsymbol{d}_{obs} \vert  \theta) \boldsymbol{\pi}_{prior}(\theta) = \exp (f(\theta))p(\theta),
\end{equation}
where $p(\theta)$ is an  auxiliary distribution that can be freely 
chosen.
According to \eqref{unnormalized posterior}, we set the target function as
\begin{equation} 
f(\theta)  = \log\left(
\frac{\boldsymbol{\tilde{\pi}}(\theta \vert \boldsymbol{d}_{obs})}{p(\theta)}
\right)\,.
\label{eq:gp_target}
\end{equation} 
In this formulation, the target function is smoothed by the logarithm, ensuring that the resulting posterior distribution, $\boldsymbol{\tilde{\pi}}(\theta \vert \boldsymbol{d}_{obs})$, remains positive. It is also important to note that if the auxiliary distribution $p(\theta)$ provides a good approximation of the posterior, then $f(\theta)$ will be nearly constant and easier to approximate. Therefore, selecting an appropriate $p(\theta)$ can simplify the approximation process and enhance the accuracy of the model.

It is clear that the optimal choice for $p(\theta)$ would be the posterior distribution. However, this is impractical in the context of inverse problems.
Therefore, starting with 
an initial density 
$p_0(\theta)$, 
an adaptive iterative framework is employed to refine $p(\theta)$, 
allowing it to converge to
the posterior distribution $\boldsymbol{\pi}(\theta \vert \boldsymbol{d}_{obs})$. 
The procedure proceeds as follows. 
Denote by $p_{n-1}(\theta)$ the current distribution at the $n$-th iteration. 
First, a Gaussian process model $\hat{f}_{n-1}(\theta)$ is constructed for $f_{n-1}(\theta)$, which is the log of the ratio between the posterior and the prior 
\begin{equation}
f_{n-1}(\theta) := \log\left(\frac{\boldsymbol{\tilde{\pi}}(\theta \vert \boldsymbol{d}_{obs})}{p_{n-1}(\theta)}\right). 
\label{f(x)} 
\end{equation}
Then, the GP surrogate $\hat{f}_{n-1}(\theta)$ is used to update $p_{n}(\theta)$ based on the approximation
\begin{equation}
    p_{n}(\theta) \propto  \exp (\hat{f}_{n-1}(\theta))p_{n-1}(\theta).
    \label{eq:app_pos}
\end{equation}
A detailed scheme can be found in Algorithm \ref{alg:adaptiveGP}.

\begin{algorithm}[H]
\caption{The adaptive GP algorithm}\label{alg:adaptiveGP}
\begin{algorithmic}[1]
\Require 
Maximum number of iterations
$N^{\max}$,
KL divergence threshold $\delta_{KL}$, maximum consecutive times $N_{KL}^{\max}$.
\Ensure The approximate posterior distribution $p_n(\theta)$.
\State{Let $\hat{p}_0(\theta) = \boldsymbol{\pi}_{prior}(\theta)$, $n_{KL}=0$.}
\State{
Select $m_0$ initial design points ${\theta_i}_{i=1}^{m_0}$, and compute $y_i = f(\theta_i)$ for $i = 1, \dots, m_0$, with $f$ defined in~\eqref{eq:gp_target}.
}
\State{Define the initial training data set as $S_0 = \{\theta_i, y_i:= f(\theta_i) \}_{i=1}^{N_{m_0}}$.}
\State{Construct an initial GP surrogate model $\hat{f}_0(\theta)$ for $f_0(\theta)$ (see \eqref{f(x)}) using the data set $S_0$.}
\For{$n = 1, \ldots, N^{\max}$}  
\State{
Draw $N_m$ samples, denoted $A_n$, from the approximate posterior $p_n(\theta) \propto \exp(\hat{f}_{n-1}(\theta)) \hat{p}_{n-1}(\theta)$ using MCMC.
}
\State{Obtain an estimated probability density function (PDF) $\hat{p}_{n}$ from samples $A_n$.} 
\State{Compute $D_{KL}(\hat{p}_{n-1}(\theta), \hat{p}_{n}(\theta))$.}
\If{$D_{KL}(\hat{p}_{n-1}(\theta), \hat{p}_{n}(\theta))< \delta_{KL}$}
\State{$n_{KL} = n_{KL}+1$.}
\Else
\State{Reset $n_{KL}=0$.}
\EndIf
\If{$n_{KL}= N_{KL}^{\max}$}
\State{Break the for loop.}
\EndIf
\State{Choose $m$ design points $\{\theta_i \}_{i=1}^{m}$ and compute $y_i = f(\theta_i)$ for $i=1,\ldots,m$ ($f$ is defined in \eqref{eq:gp_target}).}
\State{Enlarge the training data set as $S_{n} = S_{n-1} \cup \{(\theta_i, f(\theta_i)\}_{i=1}^{m}$.}
\State{Construct the GP surrogate model $\hat{f}_{n}(\theta)$ with $\hat{p}_n(\theta)$ and the training data $S_n$}
\EndFor
\end{algorithmic}
\end{algorithm}

Termination can be triggered by either of two criteria. The first criterion is reaching the maximum number of iterations. The second criterion is based on the Kullback-Leibler (KL) divergence between $p_{n-1}$ and $p_n$. If the $D_{KL}(p_{n-1}, p_n)$ is smaller than a specified threshold for $N_{KL}^{\max}$ consecutive iterations, it indicates that $p_t$ has converged.

\subsection{Iterative local updating ensemble smoother (ILUES)} \label{sec:ilues}
As mentioned earlier, GP has its own limitations. Being a kernel-based method, 
the performance of GP approximation on test data heavily depends on the quality of the training data.
In Algorithm \ref{alg:adaptiveGP}, 
the selection of design points plays a crucial role in the effectiveness of the GP.
Therefore, ILUES has been introduced as an efficient method for generating the training data. To present the ILUES method, let us first begin with the ensemble smoother. The ensemble smoother (ES) \citep{emerick2013ensemble} is a fast and efficient method for the parameter estimation in nonlinear problems. 
For an ensemble of parameters $[\theta_1^t,\ldots, \theta_{N_e}^t]$ 
at the $t$-iteration, 
where each $\theta_j^t\in \mathbb{R}^{\bar{M}}$,
the ES updates the ensemble with 
\begin{equation*}
\mathbf{\theta}_j^{t+1} = \theta_j^t + \boldsymbol{\Sigma}_{\Theta \mathbf{D}}^t( \boldsymbol{\Sigma}_{\mathbf{D} \mathbf{D}}^t + \boldsymbol{\Sigma}_{\eta})^{-1}(\tilde{\boldsymbol{d}}_j - \mathcal{G}(\theta_j^t)),
\quad j = 1,2, \cdots, N_e
\end{equation*} 
where $N_e$ denotes the number of ensemble members and 
$\boldsymbol{\Sigma}_{\eta}$ is the covariance matrix of the measurement errors.  
The empirical covariances $\boldsymbol{\Sigma}_{\Theta \mathbf{D}}^t$ and $\boldsymbol{\Sigma}_{\mathbf{D} \mathbf{D}}^t$ are given by 
	\begin{align*}
 \boldsymbol{\Sigma}_{\Theta \mathbf{D}}^t = \frac{1}{N_e-1} \sum_{i=1}^{N_e}(\theta^t_i - \Bar{\theta}^t) \otimes (\mathcal{G}(\theta^t_i) - \Bar{\mathcal{G}}^t)\quad \mbox{and}\quad 
\boldsymbol{\Sigma}_{\mathbf{D} \mathbf{D}}^t = \frac{1}{N_e-1} \sum_{i=1}^{N_e}(\mathcal{G}(\theta^t_i) - \Bar{\mathcal{G}}^t) \otimes (\mathcal{G}(\theta^t_i) - \Bar{\mathcal{G}}^t),
 	\end{align*}
respectively, where the symbol $\otimes$ denotes the Kronecker matrix product, 
$\Bar{\theta}^t$ is the average of the ensemble $\{\theta^t_j \}_{j=1}^{N_e}$, and $\Bar{\mathcal{G}}^t$ is the average of  $\{\mathcal{G}(\theta^t)_j\}_{j=1}^{N_e}$.

Note that $ \mathbf{\Sigma}_{\Theta \mathbf{D}}^t$ is the $\bar{D} \times \bar{M}$ cross-covariance matrix between the parameter samples $\{\theta^t_j\}_{j=1}^{N_e}$ and the corresponding predicted data $\{\mathcal{G}(\theta^t)_j\}_{j=1}^{N_e}$, $\boldsymbol{\Sigma}_{\mathbf{D} \mathbf{D}}^t$ is the $\bar{D} \times \bar{D}$ auto-covariance matrix of predicted data  $\{\mathcal{G}(\theta^t_j)\}_{j=1}^{N_e}$, and $\tilde{\mathbf{d}}_j$ is the vector of observed data with  $\tilde{\boldsymbol{d}}_j \sim \mathcal{N}(\boldsymbol{d}_j,\boldsymbol{\Sigma}_{\eta})$ where $\boldsymbol{d}_j = \mathcal{G}(\theta_j^t)$.
It is clear from the update scheme of the ES that the method mainly depends on the mean and covariance information.
Hence, when the posterior distribution is multimodal, ES would be unreliable. 
To address this, \citep{zhang2018iterative} proposed iterative local updating ensemble smother (ILUES) algorithm,
which extends ES to handle multimodal posterior distributions.
Although the target distribution is multimodal, the local distribution remains unimodal. By leveraging this property, ILUES employs a local update strategy.

For $\theta \in \{ \theta_j^t\}_{j = 1}^{N_e}$,  let 
\begin{equation}
J(\theta) = J_1(\theta)/ J_1^{max} + J_2(\theta)/ J_2^{max} 
\label{eq:j}
\end{equation} 
be an integrated distance measure that accounts for both the discrepancy with the observed data $\boldsymbol{d}_{\mathrm{obs}}$ and the distance to the sample $\theta_j^t$, 
where $J_1(\theta) = (\mathcal{G}(\theta) -\boldsymbol{d}_{obs})^T\boldsymbol{\Sigma}_{\eta}^{-1}(\mathcal{G}(\theta) -\boldsymbol{d}_{obs})$ evaluates the discrepancy between the forward model $\mathcal{G}(\theta)$ and data $\boldsymbol{d}_{obs}$, and $J_2(\theta) = (\theta -\theta_j^t)^T\boldsymbol{\Sigma}_{\Theta \Theta}^{-1}(\theta -\theta_j^t)$ 
quantifies the distance between
the parameters $\theta$ and sample $\theta_j^t$. Here, $\boldsymbol{\Sigma}_{\Theta \Theta}$ denotes the $\bar{M} \times \bar{M}$  auto-covariance matrix of parameters $\theta$. The normalization constants $J_1^{max}$ and $J_2^{max}$ are the maximum values of $J_1(\theta)$ and $J_2(\theta)$, respectively. 
%
The local ensemble associated with $\theta_j^t$ is then defined as the subset of $N_l$ samples with the smallest values of $J(\theta)$, where $N_l = \alpha N_e$ and $0 < \alpha < 1$. This local ensemble is denoted by $\{\theta_{j,\,i}^t\}_{i=1}^{N_l}$. The ILUES update rule can therefore be expressed as
\begin{equation} 
\label{eq localensemble}
\theta_{j,\,i}^{t+1} = \theta_{j,\,i}^t + \boldsymbol{\Sigma}_{\Theta \mathbf{D}}^{l,\,t}( \boldsymbol{\Sigma}_{\mathbf{D} \mathbf{D}}^{l,\,t} + \boldsymbol{\Sigma}_{\eta})^{-1}(\tilde{\boldsymbol{d}}_i - \mathcal{G}(\theta_{j,\,i}^t)),
\end{equation}
for $i = 1,2,\cdots,N_l$. 
Here, $ \mathbf{\Sigma}_{\Theta \mathbf{D}}^{l,\,t}$ is the $\bar{D} \times \bar{M}$ cross-covariance matrix between the local ensemble $\{\theta^t_{j,\,i}\}_{i=1}^{N_l}$ and the corresponding predicted data $\{\mathcal{G}(\theta^t_{j,\,i})\}_{i=1}^{N_l}$, while $\boldsymbol{\Sigma}_{\mathbf{D} \mathbf{D}}^{l,\,t}$ is the $\bar{D} \times \bar{D}$ auto-covariance matrix of predicted data $\{\mathcal{G}(\theta^t_{j,\,i})\}_{i=1}^{N_l}$. 
Finally, a sample is randomly selected from $\{\theta_{j,\,i}^{t+1}\}_{i=1}^{N_l}$ to serve as the updated sample $\theta_j^{t+1}$.
The ILUES  method is summarized in Algorithm \ref{alg2:ilues}. 

\begin{algorithm}[H]
\caption{Iterative local updating ensemble smoother (ILUEs)}\label{alg2:ilues}
\begin{algorithmic}[1]
\Require Forward model $\mathcal{G}(\theta)$, observational data $\boldsymbol{d}_{obs}$, maximum number of iterations $N_{iter}$, ensemble size $N_e$.
\Ensure $\{\theta_j^{N_{iter}}\}_{j=1}^{N_e}$.
\State{Sample $\{ \theta_{j}^0 \}_{j = 1}^{N_e}$ from the prior distribution $\boldsymbol{\pi}_{prior}(\theta)$  and calculate the predicted data  $\{ \mathcal{G}(\theta_{j}^0) \}_{j=1}^{N_e}$ correspondingly.} 
\For{$t = 1 \cdots N_{iter}$}
\For{$j = 1 \cdots N_e$}
\State{Calculate  $J(\theta)$ for each $\theta \in \{ \theta_{j}^t \}_{j = 1}^{N_e} $ using \eqref{eq:j}. }
\State{Calculate the updated local ensemble $\{\theta^{t+1}_{j,\,i}\}_{i=1}^{N_l}$} using \eqref{eq localensemble}.
\State{Draw the updated sample $\theta^{t+1}_j$ form $\{\theta^{t+1}_{j,\,i}\}_{i=1}^{N_l}$ randomly.} 
\EndFor
\EndFor
\end{algorithmic}
\end{algorithm}


When the ensemble size is small, the posterior distribution may not be accurately captured. However, the ensemble can rapidly concentrate in regions of high posterior probability, which is particularly useful for multimodal distributions. Therefore, in our paper, instead of using ILUES directly as a sampler for the posterior, we employ it as an efficient generator of training data. 
We will describe it in more detail in Section~\ref{sec:main_alg}.

\subsection{MCMC with the Gaussian mixture proposal}
\label{sec:mcmc_mg}
Through iterative sampling and probabilistic transitions, 
the MCMC algorithm \citep{robert2013monte} allows 
us to generate samples from the target distribution $P(\theta)$.
Starting from an 
initial point $\theta^0$, at iteration $t-1$, MCMC generates a candidate point $\theta^*$ 
from the proposal distribution $Q(\theta^{*} \vert \theta^{t-1})$. 
The candidate is then accepted with the probability 
\begin{equation*}
\alpha(\theta^{*},\theta^{t-1}) = 
\min\left\{1,\frac{P(\theta^*)}{Q(\theta^{*} \vert \theta^{t-1})} \cdot
\frac{ Q(\theta^{t-1} \vert \theta^{*}) }{P(\theta^{t-1} )} \right\}.
\end{equation*}
If $\theta^*$ is accepted, the new sample $\theta^t$ is updated as $\theta^{t} = \theta^*$; 
otherwise, the current sample is retained, i.e., $\theta^{t} = \theta^{t-1}$. 
A popular choice for the proposal distribution is the Gaussian distribution \citep{hastings1970monte}, which is symmetric, meaning that $Q(\theta^{*} \vert \theta^{t-1}) = Q(\theta^{t-1} \vert \theta^{*}) $. 
If the covariance of the proposal is small, the Markov chain randomly walks in a small parameter region. 
Conversely, if the covariance is too large, most candidate samples are likely to be rejected.

%
When the posterior distribution is multimodal, 
one can adopt a Gaussian mixture distribution as the proposal distribution when using MCMC to generate a large number of samples from the approximate posterior distribution $p_n(\theta)$~\cite{che2025stable}.
A Gaussian mixture (GM) distribution is a superposition of $K$ different
multivariate Gaussian components and is defined as
  \begin{align*}
q(\theta) = \sum_{j=1}^K w_j 
\mathcal{N}(\theta\vert\mu_j, \Sigma_j)\,,
  \end{align*}
where 
$w_j  \ge 0$ are the mixture weights satisfying $\sum_{j=1}^{K} w_j = 1$. 

Suppose that the GM proposal at iteration $t-1$ is characterized by the parameters $\{w_j^{t-1}, \mu_j^{t-1},\Sigma_j^{t-1} \}_{j=1}^{K}$. 
Then the proposal distribution at iteration $t-1$ is given by
\begin{equation} \label{proposal}
q(\theta \vert \theta^{t-1}) = \sum_{j = 1}^{K} w_j^{t-1}\mathcal{N}(\theta \vert \mu_j,\Sigma_j^{t-1}).
\end{equation}
Assume that the current sample $\theta^{t-1}$ belongs to mode $j$. Denote by $m_j^{t-1}$ the number of samples previously assigned to mode $j$. Draw a candidate sample $\theta^t$ from the proposal distribution~\eqref{proposal}. Suppose that $\theta^t$ is assigned to mode $i$. Then the GM parameters can be updated as follows:
	\begin{align}
	m_i^t      &= m_i^{t-1} +1, \,\, m_j^t = m_j^{t-1}, \quad j\neq i, \label{m_update}\\
	\mu_i^t    &=\frac{1}{m_i^t}\theta^{t} + \frac{m_i^t-1}{m_i^t}\mu_i^{t-1}, \,\,\mu_j^t = \mu_j^{t-1}, \quad j\neq i, \label{mean_update}\\
	w_j^t      &= \frac{m_j}{t}, \quad j = 1, \cdots, K, \label{weight_update}\\
	\Sigma_i^t &= \frac{1}{m^t_i}\left(\frac{(\theta^{t} - \mu_i^{t})(\theta^{t} - \mu_i^{t})^T}{m_i^t} + \epsilon \boldsymbol{I}\right) + \frac{m_i^t -2}{m_i^t-1}\Sigma_i^{t-1}, \nonumber \\
 \Sigma_j^t & = \Sigma_j^{t-1}, \quad j\neq i \label{cov_update},   
 	\end{align}
where $\epsilon > 0$ is a small number ensuring that the covariance matrix $\boldsymbol{\Sigma}_i^t$ remains positive definite~\citep{haario2001adaptive}, and $\boldsymbol{I}$ denotes the identity matrix of appropriate dimension.
For the target distribution $p_n(\theta) \propto \exp(\hat{f}_{n-1}(\theta))p_{n-1}(\theta)$, the acceptance probability is given by
\begin{equation}
   \label{accept reject}  
   \alpha(\theta^*, \theta^{t-1}) = \min\left\{
   1, \frac{\exp(\hat{f}_{n-1}(\theta^*))p_{n-1}(\theta*)}{q(\theta^*|\theta^{t-1})}\cdot \frac{q(\theta^{t-1}|\theta^{*})}{ \exp(\hat{f}_{n-1}(\theta^{t-1}))p_{n-1}(\theta^{t-1})}
   \right\}\,.
\end{equation}

To summarize, the procedure for the adaptive Metropolis MCMC with a Gaussian mixture proposal is as follows. The algorithm begins with an initial state $\theta^0$ and an initial proposal distribution defined by the initial parameters of the Gaussian mixture model. At the $t$-th step, a candidate state $\theta^*$ is first generated from the current proposal $q(\theta^* \mid \theta^{t-1})$ in~\eqref{proposal}. The candidate state is then accepted with the acceptance probability $\alpha(\theta^*, \theta^{t-1})$ in~\eqref{accept reject}). Once the new state $\theta^*$ is accepted, the parameters defining the Gaussian mixture model are updated using~\eqref{m_update}-\eqref{cov_update}. The algorithm is summarized in Algorithm~\ref{alg:mh mcmc}.

\begin{algorithm}[H]
\caption{ Adaptive Metropolis MCMC with Gaussian Mixture Proposal}
\label{alg:mh mcmc}
\begin{algorithmic}[1]
\Require Length of the Markov chain $N$, initial parameters for the Gaussian mixture distribution $\{w_k^0, \mu_k^0, \Sigma_k^0\}$ and $K$.
\Ensure Samples $\{ \theta_i\}_{i=1}^{N}$.
\State{Generate the initial state $\theta^{0}$ from the prior $\boldsymbol{\pi}_{prior}(\theta)$.} 
\For{$t = 1,\ldots, N$}
\State{Sample $\theta^{*} \sim q(\theta^{*} \vert \theta^{t-1})$ (the proposal distribution defined in~\eqref{proposal})}
\State{Compute the acceptance probability $\alpha(\theta^{*},\theta^{t-1})$ using equation \eqref{accept reject} with the proposal distribution \eqref{proposal}.
}
\State{Sample $z$ from a uniform distribution $\mathcal{U}(0,1)$.
}
\If {$\alpha(\theta^{*},\theta^{t-1}) \geq z$}
     \State{Set $\theta^{t} = \theta^{*}$.}
\Else
     \State{Set $\theta^{t} = \theta^{t-1}$.}
\EndIf
\State{Update the parameters of the Gaussian mixture proposal $ \{m_k^t, w_k^t, \mu_k^t, \Sigma_k^t\}$ for $k = 1, \cdots, K$ using~\eqref{m_update}-\eqref{cov_update}.}
\EndFor
\end{algorithmic}
\end{algorithm}

\section{Our method: ILUES-based adapative Gaussian process regression (ILUES-AGPR)}
\label{sec:main_alg}

Now, we are ready to present our method, the ILUES-based adaptive Gaussian process regression (ILUES-AGPR), to solve Bayesian inverse problems with multimodal posterior distributions. As shown in~\eqref{unnormalized posterior}, the unnormalized posterior density $\tilde{\bm{\pi}}(\theta \mid \bm{d}_{\mathrm{obs}})$ can be expressed as the product of $\exp\left(f(\theta)\right)$ and an auxiliary distribution $p(\theta)$.
Using $p(\theta)$, we construct 
a GP surrogate model $\hat{f}(\theta)$ to approximate $f(\theta)$.
Starting from an initial density $p_0(\theta)$, we then iteratively update the GP surrogate $\hat{f}_n(\theta)$ and the approximate posterior $p_n(\theta)$ using~\eqref{f(x)} and~\eqref{eq:app_pos}, respectively, until $p_n(\theta)$ converges or the maximum number of iterations is reached.

However, 
constructing surrogate models solely based on the prior distribution can be inefficient when the posterior deviates significantly from the prior. Ideally, the surrogate model should be trained using samples drawn from the posterior distribution. However, this is impractical because the posterior is not available in advance. To address this challenge, we employ the ILUES method to adaptively generate informative training data concentrated in regions of high posterior density.
In particular, when the ensemble size is small, ILUES tends to concentrate samples near the primary modes of the true posterior, even if the distribution of these points does not fully match the posterior.

\begin{algorithm}
\caption{ILUES-based Adaptive Gaussian Process Regression (ILUES-AGPR)}\label{alg3:Framework}
\begin{algorithmic}[1]
\Require{Initial number of iterations $n_0$, ensemble size $N_e$ for ILUES, forward model $\mathcal{G}(\theta)$, observational data $\bm{d}_{obs}$, maximum number of iterations $N^{\max}$ for the overall algorithm, KL divergence threshold $\delta_{KL}$, and maximum consecutive iterations $N_{KL}^{\max}$.}
\Ensure{The approximated posterior distribution $p_n(\theta)$.}

    \State{Let $n_{KL} = 0$. Draw an initial archive $\boldsymbol{Z} = \{ (\theta_j^{n}, \boldsymbol{ \tilde{\pi}}(\theta_j^{n} \vert \boldsymbol{d}_{obs})), \,j=1, \ldots, N_e, \, n = 1, \ldots, n_0\}$ using Algorithm \ref{alg2:ilues}, with $n_0$ iterations and the prior distribution $\boldsymbol{\pi}_{prior}(\theta)$. Denote by $\bm{E}_0 = \{ \theta_j^{n_0} \}_{j=1, \ldots, N_e}$ the ensemble samples.}
    
    \State{Estimate the PDF for $p_0(\theta)$ using the last ensemble samples $\bm{E}_0$, and denote it as $\hat{p}_0(\theta)$.}
    
    \State{Construct the initial training dataset $\boldsymbol{S}_0 = \{(\theta_j, f_0(\theta_j)),\,j=1, \ldots, N_e  n_0\}$ with $f_0$  defined in \eqref{f(x)}.}
    
    \State{Build the GP surrogate model $\hat{f}_0(\theta)$ using $\hat{p}_0(\theta)$ and $\bm{S}_0$.}
    
    \State{Obtain the initial parameters for the Gaussian mixture proposal using the $K$-means method: $K$ and $\{ \boldsymbol{\mu}_k, \Sigma_k \}_{k=1}^K$.}

    \For{$n = 1, \ldots, N^{\max}$}
    

          \State{Draw $M$ samples from the target density $p_n(\theta)\propto \exp(\hat{f}_{n-1})(\theta)\hat{p}_{n-1}(\theta)$ using MCMC with the Gaussian mixture proposal (see Algorithm~\ref{alg:mh mcmc}), yielding samples $A_n$.}
        

        \State{Obtain an estimated probability density function (PDF) $\hat{p}_{n}$ of $p_n(\theta)$ from samples $A_n$.}
        
        \State{Compute the KL divergence: $D_{\text{KL}}(\hat{p}_{n-1}, \hat{p}_n)$.}
        
        \If{$D_{\text{KL}}(\hat{p}_{n-1}, \hat{p}_n) \leq \delta_{\text{KL}}$}
            \State{$n_{KL} = n_{KL} + 1$.}
        \Else
            \State{Reset $n_{KL} = 0$.}
        \EndIf
        
        \If{$n_{KL} = N_{KL}^{\max}$}
            \State{Terminate the for loop.}
        \EndIf
        
        \State{Run one ILUES update iteration via~\eqref{eq localensemble} to obtain $N_e$ new design points $\{ \theta_j^{n + n_0}, \,j=1, \ldots, N_e\}$.}
        
        \State{Augment $\boldsymbol{Z}$ with 
        $\{ \theta_j^{n + n_0}, \tilde{\pi}(\theta_j^{n + n_0} \vert \boldsymbol{d}_{obs}) \}_{j=1, \ldots, N_e}$ and update the training dataset to $\boldsymbol{S}_{n+1}=\{(\theta_j^{n+n0}, f_n(\theta_j^{n+n0})),\, j=1,\ldots,N_e\}$.}
        
        \State{Construct the updated GP surrogate model $\hat{f}_n(\theta)$ with $\hat{p}_n(\theta)$ and the training dataset $\boldsymbol{S}_{n+1}$.}
        
        \State{Obtain the updated parameters for the GM proposal using the K-means method: $K$ and $\{ \boldsymbol{\mu}_k, \Sigma_k \}_{k=1}^K$.}
        
    \EndFor
\end{algorithmic}
\end{algorithm}

Specifically, we run the ILUES algorithm shown in Algorithm~\ref{alg2:ilues} with $N_e$ ensemble members for $n_0$ iterations, producing $N_e \times n_0$ parameter–model evaluation pairs $\boldsymbol{Z} = \{ (\theta_j^{n}, \boldsymbol{ \tilde{\pi}}(\theta_j^{n}\vert \boldsymbol{d}_{obs})), \,j=1,\ldots,N_{e}, \, n =1,\ldots,n_0\}$.
We then approximate 
the initial auxiliary distribution $p_0(\theta)$ using the last ensemble $\bm{E}_0 = \{\theta_j^{n_0}\}_{j=1}^{N_e}$ and denote the estimated probability density function as $\hat{p}_0(\theta)$. The initial training dataset for GP modeling is
$\boldsymbol{S}_0  = \{(\theta_j^n,  f_0(\theta_j^n)), j=1,\ldots,N_{e}, \,n = 1,\ldots,n_0\}$, where $f_0(\theta)$ is defined in~\eqref{f(x)}. Next, we construct the initial GP surrogate $\hat{f}_0(\theta)$ using $\boldsymbol{S}_0$ and $\hat{p}_0(\theta)$. Meanwhile, we apply $K$-means clustering to $\bm{E}_0$ to obtain the initial
Gaussian mixture proposal, which is fully characterized by parameters
$\{w_j, \,\mu_j,\, \Sigma_j \}_{j=1}^K$.

At iteration $n\ge 1$,
given the current prior distribution $\hat{p}_{n-1}(\theta)$ and GP surrogate $\hat{f}_{n-1}(\theta)$, we draw $M$ samples from the target density $p_n(\theta)\propto \exp(\hat{f}_{n-1})(\theta)\hat{p}_{n-1}(\theta)$ using MCMC with the GM proposal 
yielding samples $A_n$. We then approximate $p_n(\theta)$ by $\hat{p}_n(\theta)$ using  the samples $A_n$ and kernel density estimation. Next, we check the termination condition. We compute the KL divergence between the current PDF $\hat{p}_{n}$ and the previous $\hat{p}_{n-1}$. If the divergence is smaller than the threshold 
$\delta_{\text{KL}}$ in $N_{KL}^{\max}$ consecutive iterations
the algorithm terminates. Otherwise, we perform one ILUES iteration to obtain $\bm{E}_{n+1}=\{\theta_j^{n+n_0}\}_{j=1,\ldots,N_e}$. Finally, we augment $\bm{Z}$ with $\{ (\theta_j^{n+n0}, \tilde{\bm{\pi}}(\theta_j^{n+n0}|\bm{d}_{obs})),\,j=1,\ldots,N_e\}$, and update the training dataset to
  \begin{align*}
      \bm{S}_{n+1}=\{(\theta_j^{n+n0}, f_n(\theta_j^{n+n0})),\, j=1,\ldots,N_e\}
  \end{align*}
Our method is summarized in Algorithm~\ref{alg3:Framework}.

\section{Numerical examples}
\label{sec 5}

In this section, we will evaluate the performance of our proposed algorithm using two examples.
All numerical experiments were conducted using MATLAB R2024a on a Windows 11 workstation with a 12-core CPU.

\subsection{Example 1: contamination source identification}
\label{sec:ex_1}
Consider the contaminant source inversion problem in a two-dimensional domain, 
where the goal is to infer the contaminant source based on a set of observations. The forward model is described by a parabolic equation:
%
\begin{equation}
\label{eq:exam1}
\begin{aligned}
    & \frac{\partial u(\boldsymbol{x},t)}{\partial t} = D \Delta u(\boldsymbol{x},t),  \quad \boldsymbol{x} \in \Omega=[-1,1]^2, \quad t \in [0,\infty),    \\
  & u(\boldsymbol{x},t) = 0, 
  \quad \boldsymbol{x} \in \partial \Omega,   \\
 & u(\boldsymbol{x},0) = \frac{M}{2\pi h^2} \exp \left(\frac{-\|\boldsymbol{\xi} - \boldsymbol{x}\|^2}{2h^2}\right),  
\end{aligned}
\end{equation}
where $D$ is the diffusion coefficient, 
$M$ is the total amount of released contaminant, and $h$ is the radius of concentration source. 
We assume that $D$, $M$, and $h$ are known parameters and are set to be $1$, $15$, and $0.1$, respectively.
The field $u(\boldsymbol{x},t)$ represents the concentration of the contaminant, and $\bxi$ denotes the location of the contaminant.

\begin{figure}[h]%
\centering
\includegraphics[scale=0.8]{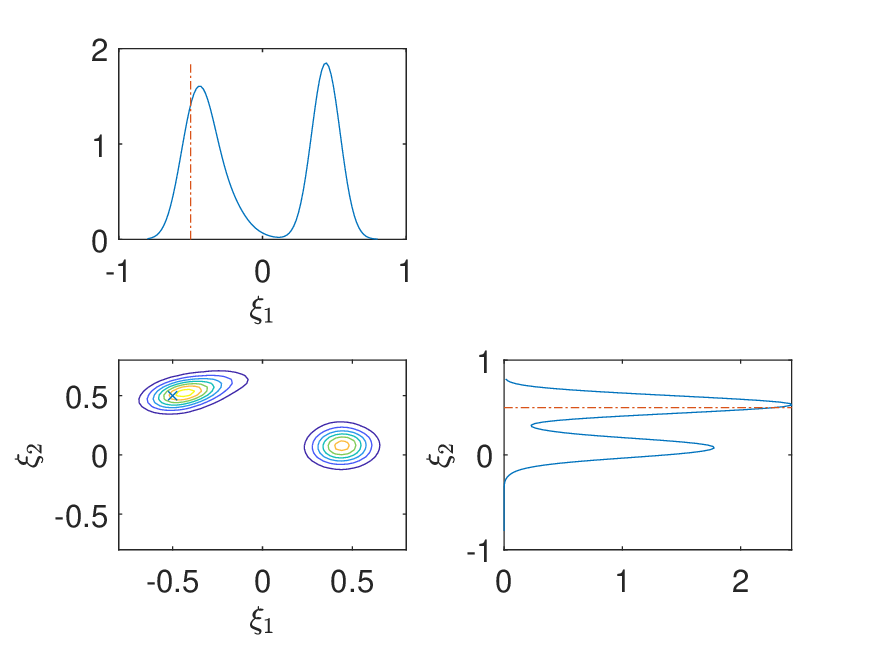}
\caption{Joint posterior distribution for Example 1 and its marginal 
probability density functions.}
\label{fig:example1-dream}
\end{figure}

In this example, the location of source, $\bxi$, is unknown and is treated as the inference parameter of interest. 
We set the prior as 
a uniform distribution $\boldsymbol{\xi} \sim \mathcal{U}([-1,1]^2) $. 
The true value is set to $\bxi=(\xi_1, \xi_2) = (-0.5, 0.5)$.
The observational data, $\bm{d}_{obs}$, are collected at the locations $(-0.4,-0.4)$ and $(0,0.4)$ at time $t = 0.04$.
We assume that the observations are contaminated by zero-mean Gaussian noise, whose standard deviation is set to 5\% of the observed value.  We use the finite difference method to solve the forward model \eqref{eq:exam1} with a uniform mesh, where $\Delta x= 0.025$ and $\Delta t = 1.25\times 10^{-4}$. 
The analytical form of the posterior distribution is not available. 
Therefore, we use the DREAM-KZS \cite{zhang2020improving}  algorithm with $8$ parallel Markov chains with $2\times 10^5$ iterations to generate the 
\textit{true} posterior distribution. 
The DREAM-KZS implementation is available at \url{https://github.com/zirandaode/DREAM_KZS}.
The posterior distribution obtained by DREAM-KZS 
for this example and the corresponding marginal probability density functions are shown in Figure~\ref{fig:example1-dream}. As seen, the distribution is multimodal. 
One mode corresponds to the true solution, while the other mode remains unknown.

\begin{figure}[!htp]
    \centerline{
    \begin{tabular}{ccc}
    \includegraphics[width=0.3\textwidth]{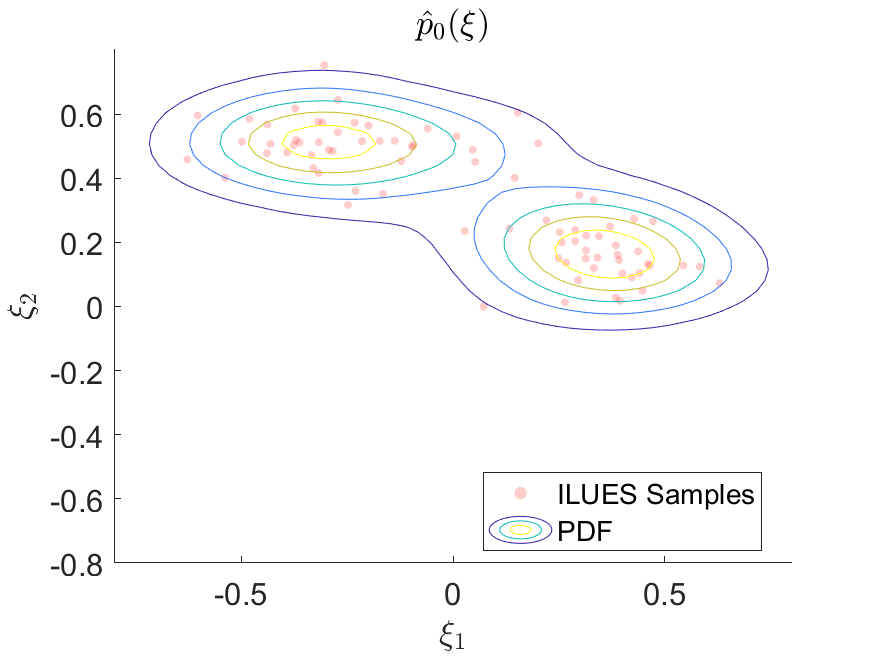}  & 
     \includegraphics[width=0.3\textwidth]{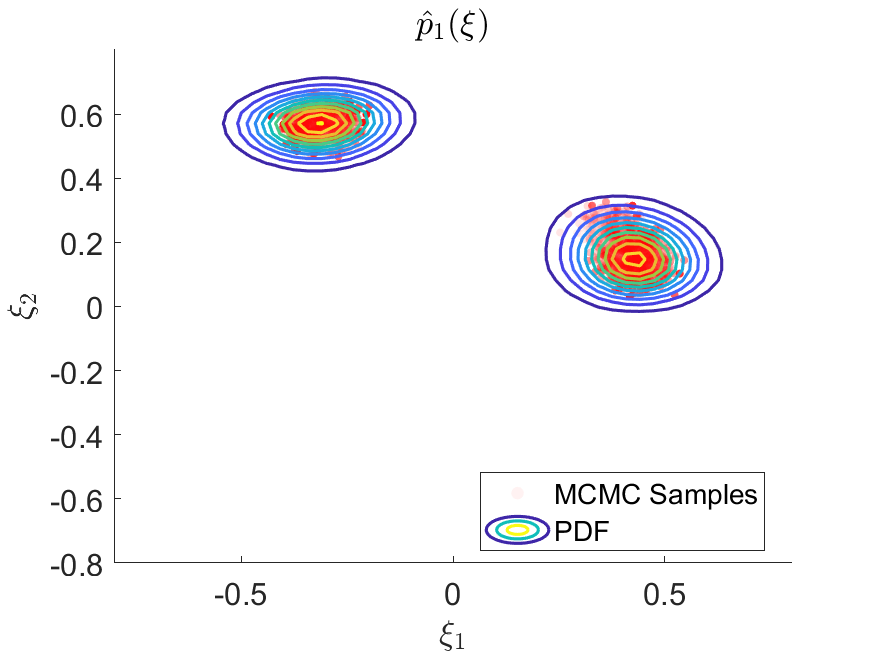} &
    \includegraphics[width=0.3\textwidth]{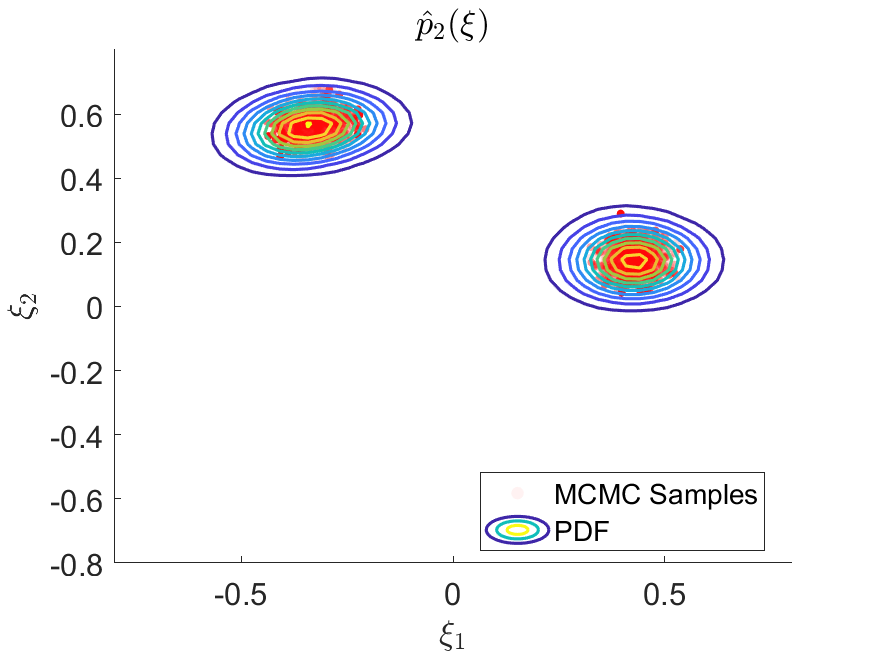}  \\
         (a) Estimated PDF $\hat{p}_0(\boldsymbol{\xi})$.  & 
     (b) Estimated PDF $\hat{p}_1(\boldsymbol{\xi})$ & 
     (c) Estimated PDF $\hat{p}_2(\boldsymbol{\xi})$ 
     \\
     \includegraphics[width=0.3\textwidth]{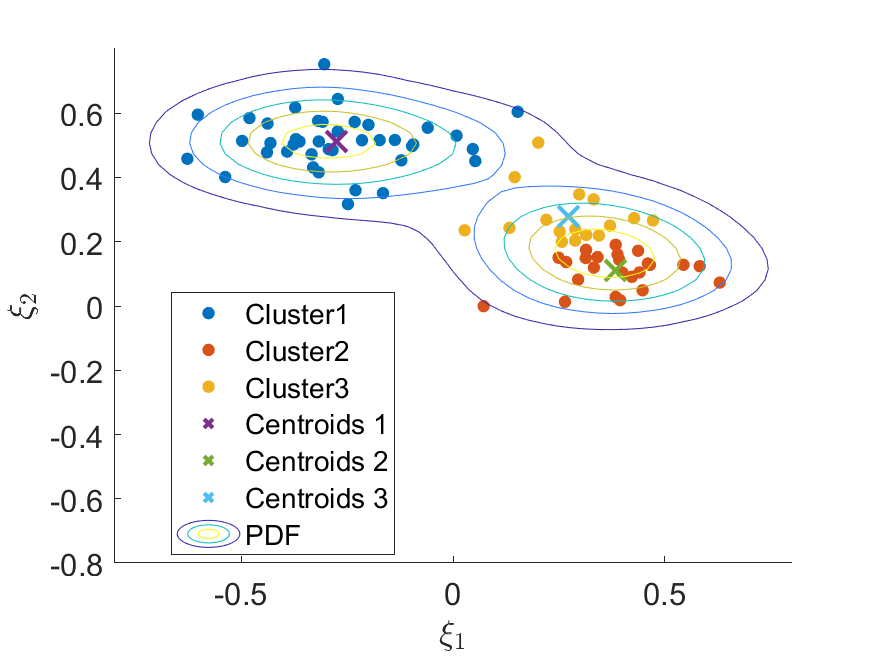}  & 
     \includegraphics[width=0.3\textwidth]{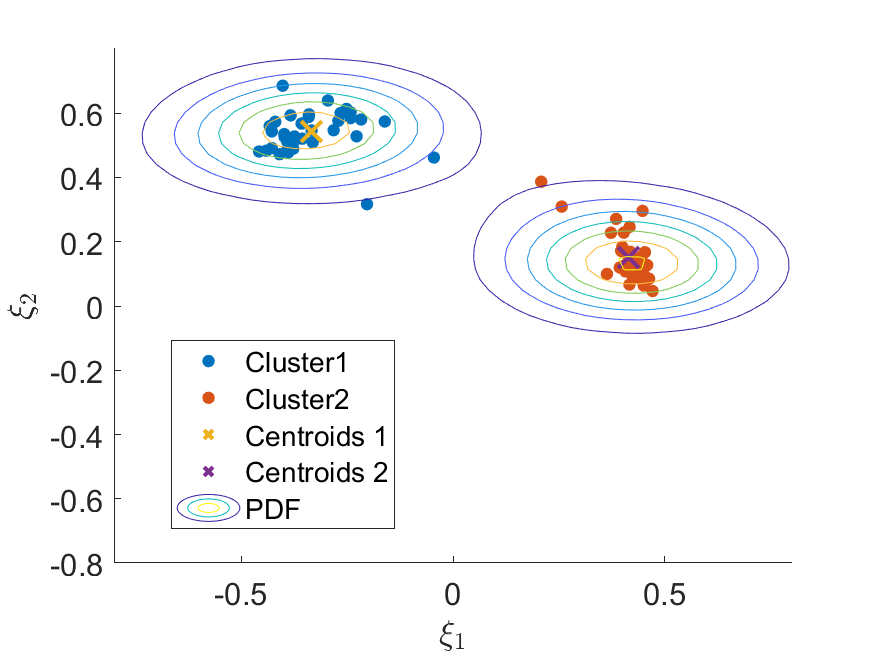} &
    \includegraphics[width=0.3\textwidth]{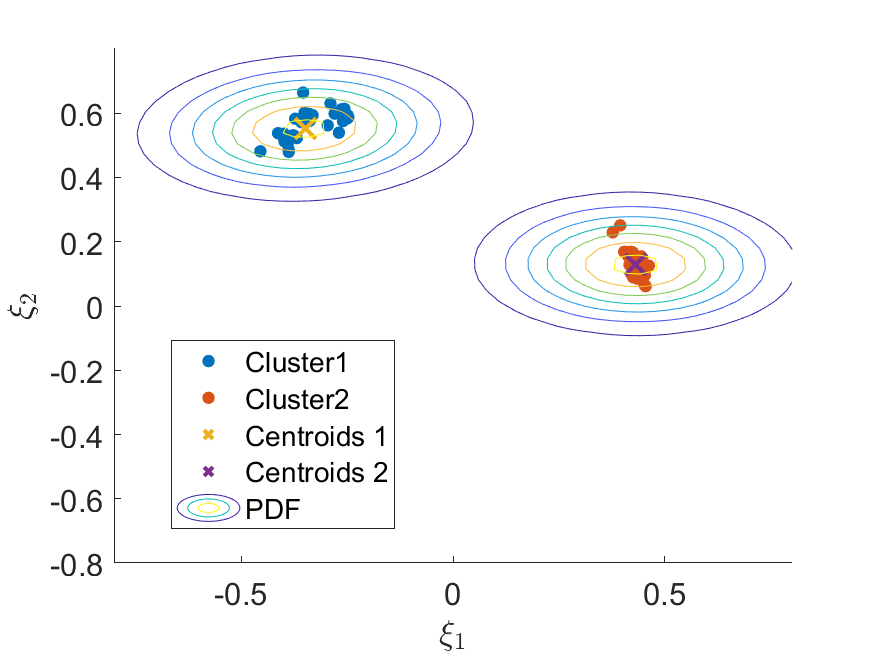} \\
    (d)  Ensemble samples at iteration $n=0$ & 
     (e)  Ensemble samples at iteration $n=1$ & 
     (f) Ensemble samples at iteration $n=2$
    \end{tabular}}
    \caption{\textbf{Top row:} Approximated posterior distributions at iterations $n=0$, $1$, and $2$. Blue dots in (a) represent ILUES ensemble points, while (b) and (c) show MCMC-generated samples. \textbf{Bottom row}: Ensemble samples at iterations $n=0$, $1$, and $2$. Dots in different colors represent distinct clusters, X marks indicate cluster centroids, and contours display the estimated PDF based on the ensemble points.}
    \label{fig:exam1-hatp-change}
\end{figure}

We apply the proposed ILUES-AGPR method to this example and compare it with other approaches. Specifically, we use the ILUES algorithm to generate the initial archive 
$\boldsymbol{Z} = \{(\boldsymbol{\xi}_j^t,  \boldsymbol{ \tilde{\pi}}(\boldsymbol{\xi}_j^{t}\vert \boldsymbol{d}_{obs})),\,j=1,\ldots,N_{e}, \,t =1,\ldots,t_0\}$,
where the forward model is defined in~\eqref{eq:exam1}, the ensemble size is set to $N_e = 80$, the number of initial iterations is $n_0 = 1$, and a uniform prior distribution is assumed. 
Using the ensemble samples, we perform kernel density estimation with the MATLAB package \textbf{kde} to approximate the initial PDF, $\hat{p}_0(\bm{\xi})$. The package is available at \url{https://ics.uci.edu/~ihler/code/kde.html}.
Next, we apply the $K$-means method \cite{kmeans_opt} to cluster the ensemble data points. Figure~\ref{fig:exam1-hatp-change}(a) shows the generated samples and the corresponding estimated PDF. Compared with the true posterior distribution shown in Figure~\ref{fig:example1-dream}, we observe that even with only one iteration, the ILUES algorithm successfully identified regions of high posterior probability, providing a reliable initial approximation for $p_0(\boldsymbol{\xi})$. Figure~\ref{fig:exam1-hatp-change}(d) presents the clustering results, where 
dots in different colors represent distinct clusters,
asterisks indicate the cluster means, and contours represent the estimated PDF based on the ensemble samples. The kernel function used in the GP regression is the squared exponential kernel, and its hyperparameters are automatically optimized using a gradient descent method.

In this example, we set the MCMC chain length ($M$ in Algorithm~\ref{alg3:Framework}) to $1 \times 10^4$ and use a burn-in rate of $20\%$. The adaptive sampling process is illustrated in panels (a)-(c) of Figure~\ref{fig:exam1-hatp-change}. The training data points 
shown in Figure~\ref{fig:exam1-hatp-change}(a) are sampled using ILUES with a single iteration. Panels (b) and (c) of Figure~\ref{fig:exam1-hatp-change} show the estimated probability density functions $\hat{p}_1(\boldsymbol{\xi})$ and $\hat{p}_2(\boldsymbol{\xi})$ obtained from the MCMC samples. The corresponding acceptance rates are $48.06\%$ and $48.60\%$, respectively, which fall within a desirable range.
In addition, the KL divergence between $\hat{p}_0(\bxi)$ and $\hat{p}_1(\bxi)$ is 0.7232, while that between $\hat{p}_1(\bxi)$ and $\hat{p}_2(\bxi)$ is 0.0422, indicating that the proposed algorithm can converge rapidly.
Panels (e) and (f) of Figure~\ref{fig:exam1-hatp-change} present the ensemble samples with clustering information, where contours denote the estimated probability density functions based on the ensemble samples at iterations $n = 1$ and $n = 2$, respectively. Compared with panels (b) and (c) of Figure~\ref{fig:exam1-hatp-change}, these results show that ILUES is capable of generating samples concentrated in regions of high posterior probability. However, when the ensemble size is small, the resulting probability density function estimates may be unreliable.

\begin{figure}[h]%
\centerline{
    \begin{tabular}{ccc}
    \includegraphics[width=0.34\textwidth]{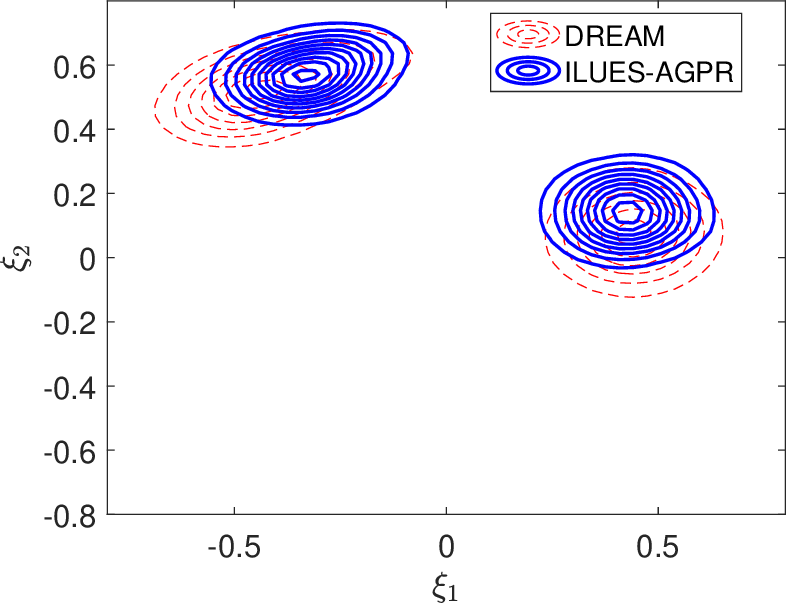}  & 
    \includegraphics[width=0.34\textwidth]{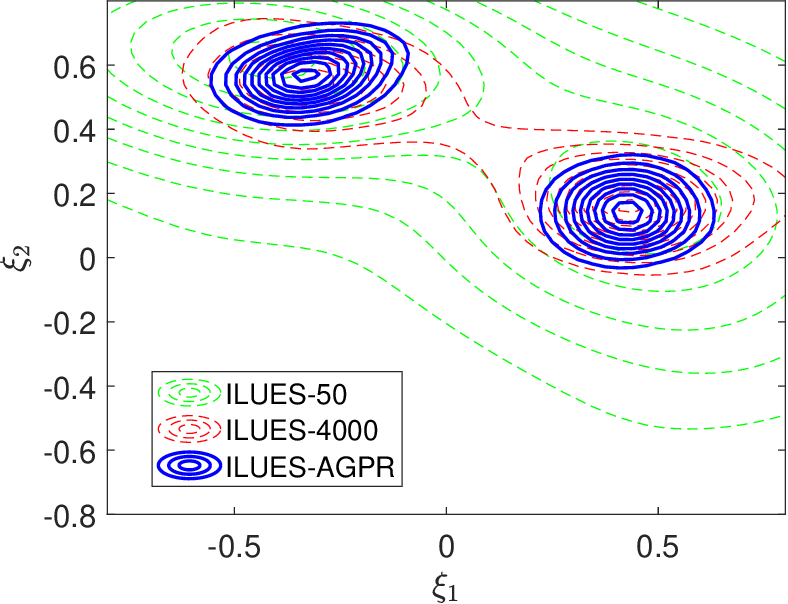} \\
    (a) & (b) 
    \end{tabular}
    }
\caption{(a) Posterior probability density functions estimated by DREAM and ILUES-AGPR. (b) Posterior probability density functions estimated by ILUES-AGPR and ILUES with two ensemble sizes, 
$N_e=50$ (ILUES-50) and $N_e=4000$ (ILUES-4000).}  
\label{fig:compare_ex1}
\end{figure}

We compare the posterior probability density functions obtained using DREAM, ILUES, and ILUES-AGPR.
Specifically, Figure~\ref{fig:compare_ex1}(a) compares the posterior probability density functions produced by DREAM and the proposed ILUES-AGPR method. 
As shown in Fig.~\ref{fig:compare_ex1}(a), ILUES-AGPR successfully captures the same two dominant posterior modes as DREAM, with comparable locations and shapes.
This indicates that ILUES-AGPR is able to recover the intrinsic multimodal structure of the posterior distribution. 
Figure~\ref{fig:compare_ex1}(b) presents the results obtained using ILUES with a single iteration for two ensemble sizes, $N_e = 50$ and $N_e = 4000$. 
Figure~\ref{fig:compare_ex1}(b) further highlights the robustness of ILUES-AGPR with respect to ensemble size.
When a small ensemble ($N_e=50$) is used, ILUES produces severely distorted posterior distributions, indicating sample degeneracy and inadequate exploration of the parameter space.
Increasing the ensemble size to $N_e=4000$ still does not lead to a satisfactory posterior approximation, suggesting that simply enlarging the ensemble is insufficient.
In contrast, ILUES-AGPR yields a stable and well-resolved multimodal posterior even with a moderate ensemble size, demonstrating its effectiveness in mitigating ensemble collapse and capturing the intrinsic multimodality of the inverse problem.

\begin{table}[h]
    \centering
    \begin{tabular}{lccc}
    \toprule 
    Method & Mean& Median & Standard Deviation\\
    \midrule 
      ILUES-50 & 0.24 & 0.23 & 0.02\\
      ILUES-4000 & 31.49 & 31.50 & 0.57 \\
      ILUES-AGPR & 45.28 & 45.18 & 0.61 \\
      DREAM & 425.53 & 426.02 & 19.16\\
      \bottomrule
    \end{tabular}
    \caption{Comparison of the running time (in seconds) for four methods.  
    }
    \label{tab:running_time}
\end{table}

Furthermore, Table~\ref{tab:running_time} compares the computational efficiency of the four methods by reporting their mean and median running times (in seconds) based on 100 independent runs. The results show that
ILUES is the fastest, ILUES-AGPR is marginally slower than ILUES-4000, and DREAM 
is the most computationally expensive. However, as illustrated in Fig.~\ref{fig:compare_ex1}(b), the fastest method, ILUES-50, yields unsatisfactory estimation accuracy for this problem. Increasing the ensemble size to 4000 (ILUES-4000) does not lead to substantial performance gains, indicating limited accuracy improvements from simply enlarging the ensemble. In contrast, ILUES-AGPR achieves significantly better estimation accuracy than ILUES-4000 while maintaining a comparable computational cost. DREAM provides more reliable posterior estimates due to its Markov chain Monte Carlo sampling mechanism, it is nearly ten times slower than ILUES-AGPR (425.53 vs. 45.28 seconds). This highlights ILUES-AGPR as the most favorable trade-off between efficiency and accuracy. These findings establish our proposed method, ILUES-AGPR, as the most favorable trade-off between computational efficiency and estimation accuracy.

\subsection{Example 2: contamination source location and strength identification}

In the second example, we consider a Poisson equation with mixed boundary conditions:
\begin{equation}
\label{eq:forward_exam2}
\begin{aligned}
    &-\nabla \cdot \bigl(a(\boldsymbol{x}) \nabla u(\boldsymbol{x})\bigr)
    = f(\boldsymbol{x}),  
    \qquad \boldsymbol{x} \in D = [0,1]^2, \\
    & u(0,y) = 0, \quad u(1,y) = 0,
\end{aligned}
\end{equation}
where the remaining two boundaries are subject to no-flow (Neumann) boundary conditions. The coefficient 
$a(\boldsymbol{x})$ denotes the permeability field, which is assumed to be uniform, i.e., 
$a(\boldsymbol{x}) = 0.2$. 
The source term is given by
\begin{equation}
\label{eq:source}
    f(\boldsymbol{x}) = \frac{\vert s \vert}{2\pi h^2} \exp\left(- \frac{\| \boldsymbol{x} - \boldsymbol{\xi} \|^2}{2h^2}\right)\,,
\end{equation}
where  $\bxi=(\xi_1, \xi_2)$ is the location of the source, $h$ and $s$ represent the width and strength of the source, respectively.
In this example, we assume that the source width $h$ is known and fixed at 0.05. Both the source strength $s$ and the source location $\boldsymbol{\xi}$ are treated as unknown parameters of interest. Accordingly, the unknown parameter vector is defined as $\boldsymbol{\theta}= (\xi_1, \xi_2, s)$.


Sensor measurements are defined through an observation operator $\mathcal{O}$ acting on the PDE solution, i.e, 
\[
\bm{d}_{obs} = \mathcal{O}(u(\bm{x};\bm{\theta})) + \eta\,,
\]
where $\mathcal{O}(u) = [u(\bm{x}^{(1)}),u(\bm{x}^{(2)}),\ldots,u(\bm{x}^{(\overline{D})})]$ corresponds to pointwise sampling of the solution at $\overline{D}$ sensor locations $\{\bm{x}……{(i)}\}_{i=1}^{\overline{D}}$.
In this example, 
the synthetic PDE solution $u$ is generated by solving the forward model~\eqref{eq:forward_exam2} on a uniform $32 \times 32$ mesh using a finite element solver~\cite{elman14finite},
the sensors are distributed on a uniform 
$3 \times 3$ grid covering the region $[0.3,0.7] \times [0.3,0.7]$. 
The true source location and strength are set to $\boldsymbol{\xi} = (0.6, 0.6)$ and $s = 1$, respectively. Measurement noise is assumed to be independent distributed Gaussian noise with zero mean and a standard deviation equal to $5\%$ of the observed value. The prior distribution of $s$ is taken as uniform on $\mathcal{U}(-2, 2)$, while the prior for the source location $\boldsymbol{\xi}$ is assumed to be uniform over the domain $D$.

\begin{figure}[h]%
\centering
\includegraphics[scale=0.9]{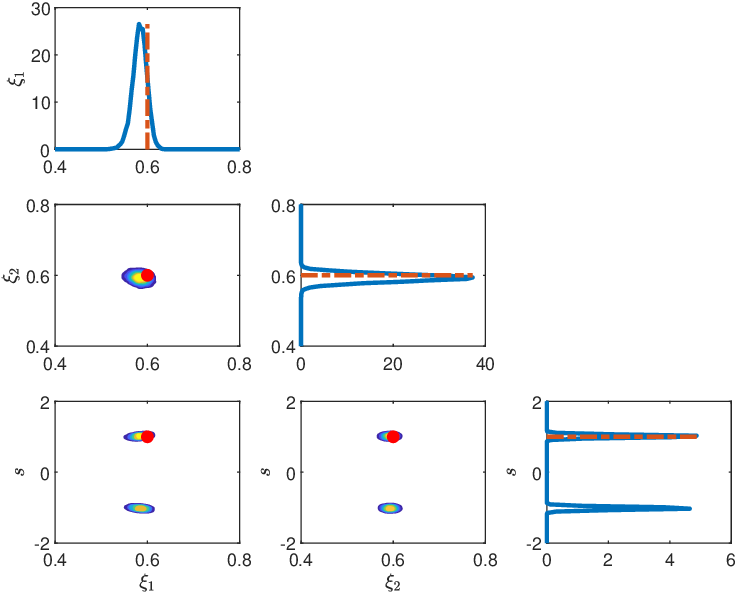}
\caption{One-dimensional (1D) and two-dimensional (2D) marginal posterior probability density functions obtained using DREAM-KZS for Example 2.}
\label{fig:example2_dream}
\end{figure}

The source function in \eqref{eq:source} is symmetric with respect to $s$, leading to a bimodal posterior distribution for $s$, with modes centered at $s=1$ and $s=-1$. In contrast, the marginal posterior distribution of the source location $\bxi$ is unimodal. To generate the \textit{true} posterior distribution, we employ the DREAM-KZS algorithm with eight parallel Markov chains and $2 \times 10^5$ iterations.
The corresponding one-dimensional (1D) and two-dimensional (2D) marginal probability density functions are shown in Figure~\ref{fig:example2_dream}.

%


We use this example to illustrate the effect of the ensemble size $N_e$. In the ILUES step of our method ILUES-AGPR, we consider two cases with $N_e = 80$ and $N_e = 150$, respectively, for which the forward model is evaluated separately at each iteration. In the following subsection, these two settings are referred to as \Neless{} and \Nemore{}, respectively. In both cases, the number of initial iterations is set to $n_0 = 4$.

\begin{figure}[h]%
\centerline{
    \begin{tabular}{ccc}
    \includegraphics[width=0.3\textwidth]{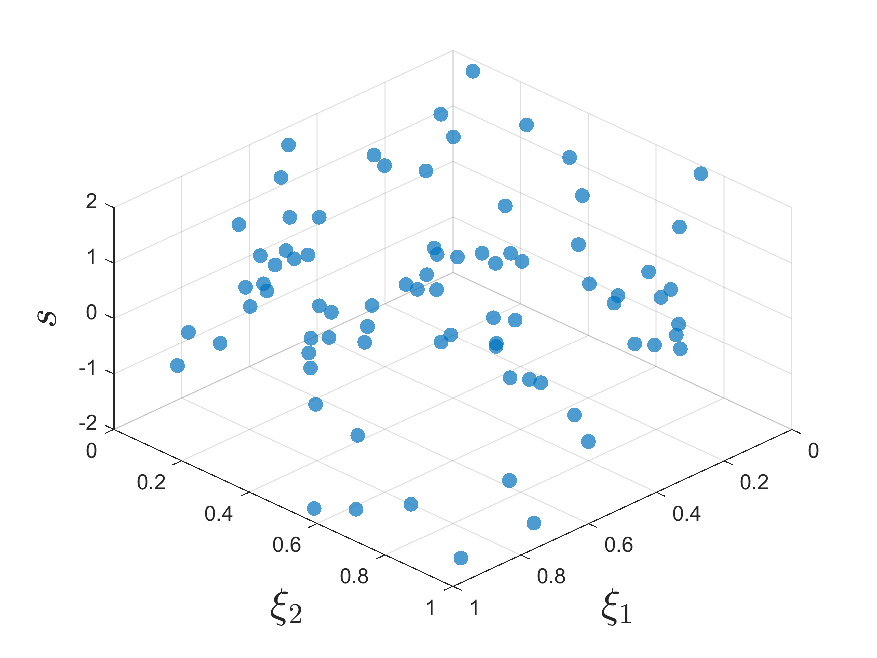}  & 
    \includegraphics[width=0.3\textwidth]{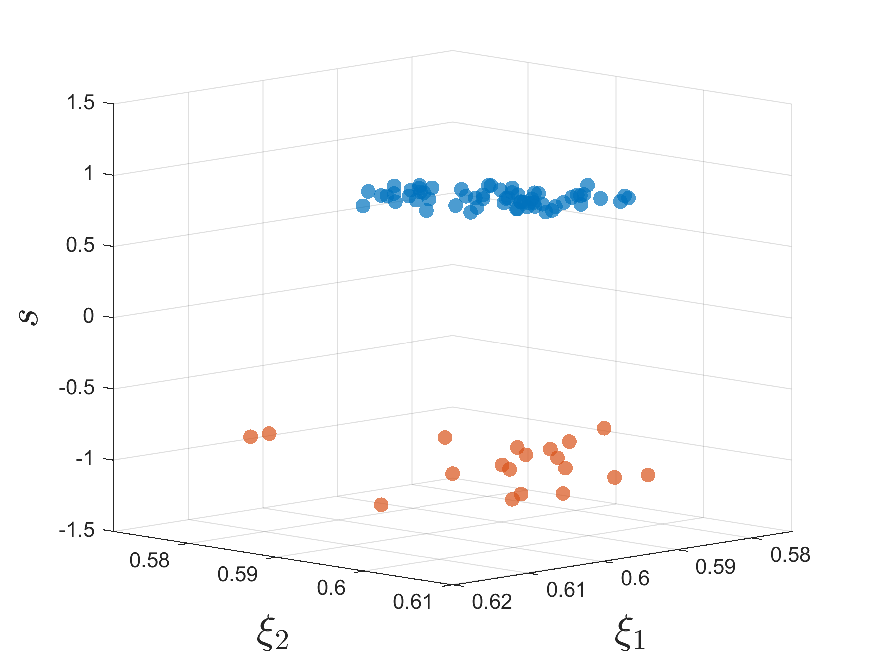} 
    &  \includegraphics[width=0.3\textwidth]{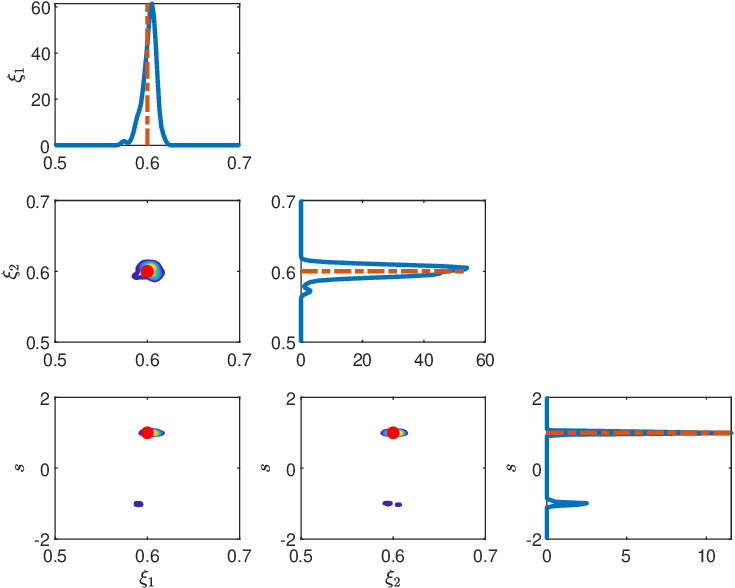} \\
    (a) Prior samples (\Neless{}) & (b) Ensemble sample at iteration $n=0$ (\Neless{}) & (c) Estimated PDF $\hat{p}_0$  (\Neless{})  \\
    & \\
    \includegraphics[width=0.3\textwidth]{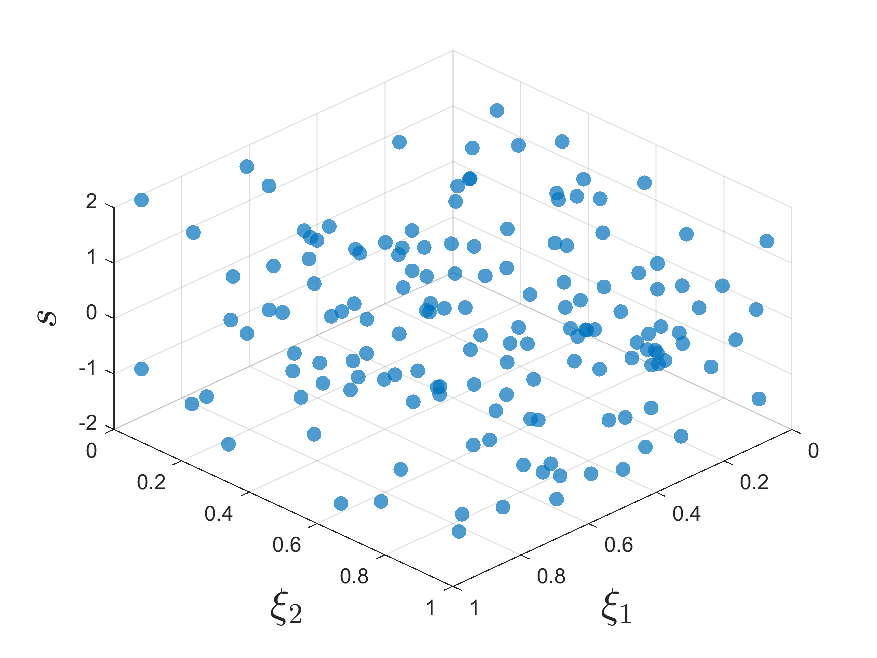}  & 
    \includegraphics[width=0.3\textwidth]{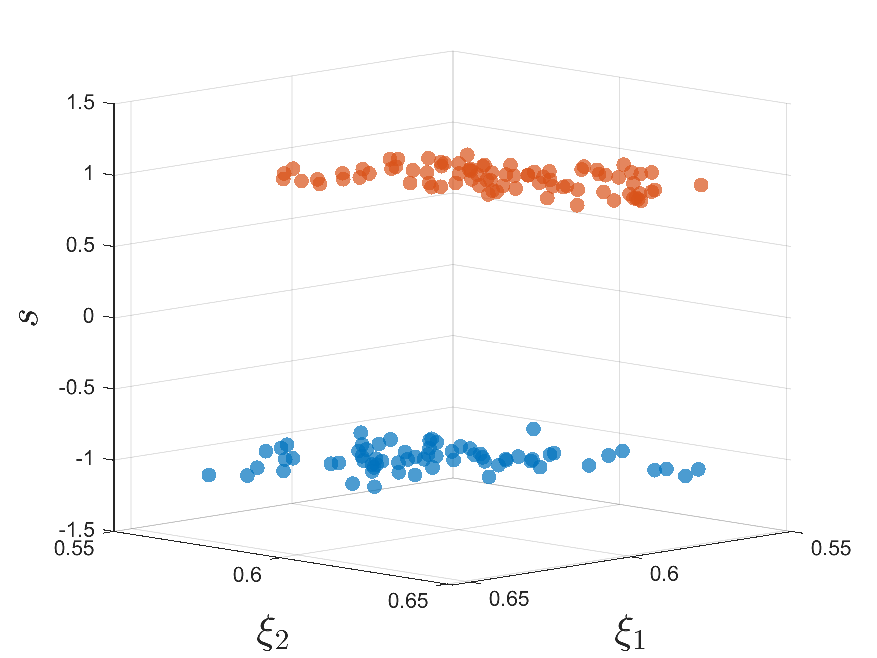} &
    \includegraphics[width=0.3\textwidth]{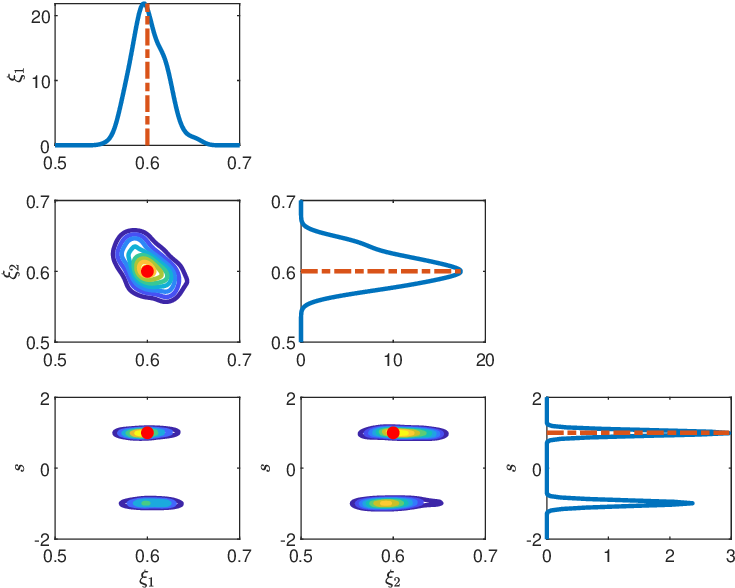}\\
    (d) Prior samples (\Nemore{}) & (e) Ensemble sample at iteration $n=0$ (\Nemore{})  & (f) Estimated PDF $\hat{p}_0$  (\Nemore{})\\
    \end{tabular}
    }
\caption{\textbf{Top row}: ILUES step for the \Neless{} case. (a) Prior samples (blue dots). (b) Ensemble samples after $n_0$ iterations with clustering, where different colors indicate different clusters. (c) Corresponding 1D and 2D marginal estimated probability density function (PDF), $\hat{p}_0$.
\textbf{Bottom row}: ILUES step for the \Nemore{} case. (d) Prior samples (blue dots). (e) Ensemble samples after $n_0$ iterations with clustering, with colors representing clusters. (f) Corresponding 1D and 2D marginal estimated PDF, $\hat{p}_0$.}
\label{fig:compare_ex2}
\end{figure}

For both the \Neless{} and \Nemore{} cases, the prior samples are uniformly distributed over the parameter space, as shown in panels (a) and (d) of Figure~\ref{fig:compare_ex2}, respectively. After $n_0$ iterations and clustering, the resulting samples are shown in panels (b) and (e), where different colors denote different clusters. The corresponding approximate distributions are presented in panels (c) and (f), respectively. These results indicate that, although the ILUES step captures both modes in each case, the distribution in the \Neless{} case tends to collapse toward a single mode.


In this example, we use the same Gaussian process settings as in Example 1. We set the MCMC chain length ($M$ in Algorithm~\ref{alg3:Framework}) to $4 \times 10^4$ and use a burn-in rate of $30\%$. 
The maximum number of iterations $N^{\max}$ is set to 6, the KL-divergence threshold $\delta_{\mathrm{KL}}$ is set to 0.05, and the maximum number of consecutive iterations $N^{\max}_{\mathrm{KL}}$ is set to 2. In the \Neless{} case, the algorithm terminates at $n = 3$. The KL divergences $D_{\mathrm{KL}}(\hat{p}_{n-1}, \hat{p}_n)$ for $n = 0, 1, 2, 3$ are 0.45, 0.16, 0, and 0, respectively. 
The corresponding MCMC acceptance rates for $n = 1, 2, 3$ are $26.70\%$, $28.61\%$, and $26.70\%$. 
In the \Nemore{} case, the algorithm also terminates at $n = 3$. The KL divergences for $n = 0, 1, 2, 3$ are 0.37, 0.19, 0.04, and 0.012, respectively, 
with MCMC acceptance rates of $27.90\%$, $24.29\%$, and $25.20\%$ for $n = 1, 2, 3$, respectively.
Overall, the convergence behavior and MCMC performance are nearly identical in both cases. However, the computational cost differs: the total runtime is 103 seconds for the \Neless{} case and 125 seconds for the \Nemore{} case. Owing to the smaller ensemble size, the \Neless{} configuration is slightly more efficient.

%

\begin{figure}
\centerline{
    \begin{tabular}{ccc}
    \includegraphics[width=0.49\textwidth]
    {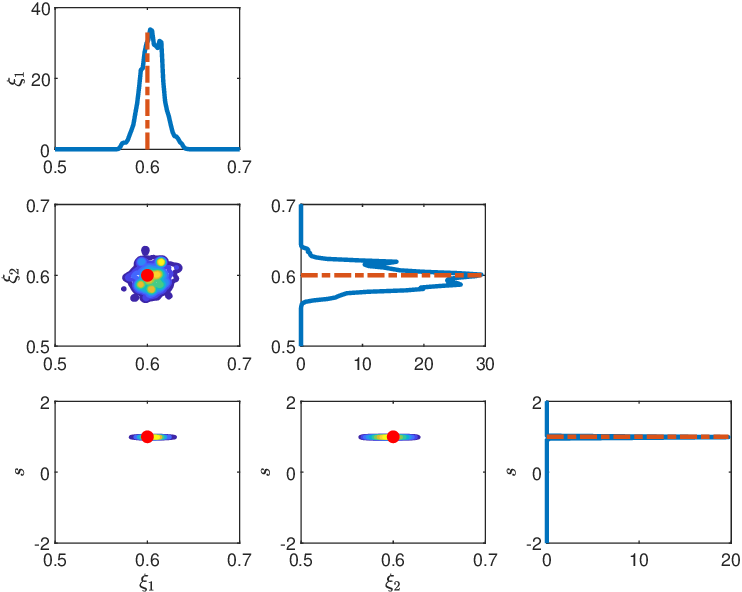}  & 
    \includegraphics[width=0.49\textwidth]{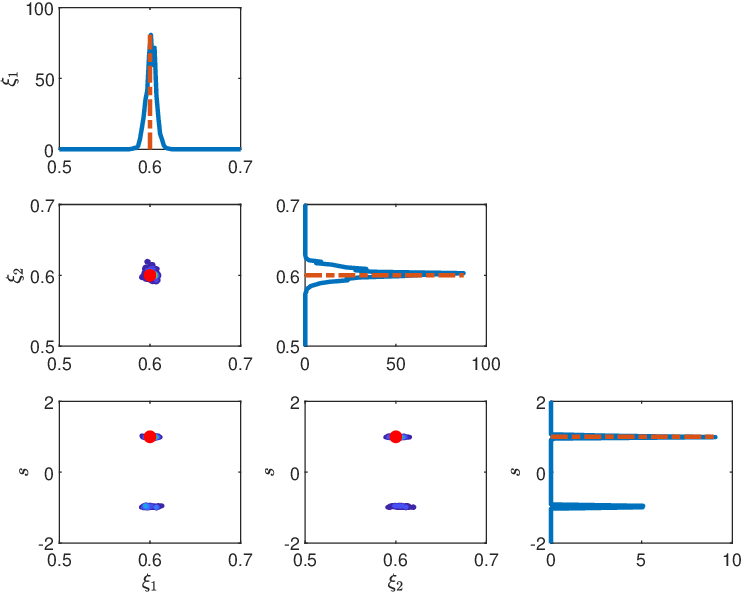} \\
    (a) & (b) \\
    \end{tabular}
    }
\caption{Estimated 1D and 2D marginal posterior probability density functions for the \Neless{} and \Nemore{} cases. (a): \Neless{}; (b): \Nemore{} case.}

\label{fig:results_ex2}
\end{figure}

Figure~\ref{fig:results_ex2} shows the marginal posterior density functions of $\theta$ along with the contours of the corresponding 2-dimensional marginal PDF. In Panel (a) of Figure~\ref{fig:results_ex2}, the final posterior for the \Neless{} case is trapped in a single mode, failing to capture the multimodal distribution. In contrast, Panel (b) shows that the \Nemore{} case successfully captures both modes. 

These results highlight the importance of choosing an appropriate ensemble size $N_e$. A small $N_e$ can lead to mode collapse, whereas a large $N_e$ may reduce computational efficiency. The impact of $N_e$ on the overall algorithm primarily arises in two aspects. First, the quality of the prior distribution generated by ILUES can vary significantly. Second, the number of training points available for the GP surrogate changes accordingly. With a smaller training dataset, the GP surrogate may perform suboptimally, resulting in \textit{jitter} in the posterior distribution generated by MCMC, as observed in Panel (a) of Figure~\ref{fig:results_ex2}.


\section{Conclusion}
\label{sec:con}


In this paper, we address the challenge of efficiently approximating multimodal posterior distributions arising in Bayesian inverse problems. We reviewed the formulation of Bayesian inverse problems and relevant existing methodologies, including adaptive Gaussian process surrogates and the ILUES method. Building on these foundations, we proposed a new approach for efficiently sampling from multimodal posterior distributions. Numerical examples demonstrate that the effectiveness and overall efficiency of the proposed ILUES-AGPR algorithm. 
\bigskip

\bibliography{references}
\end{document}